\documentstyle[12pt,epsfig]{article}

\def\thefootnote{\arabic{footnote}}
\makeatletter

\def\secteqno{\@addtoreset{equation}{section}%
\def\theequation{\thesection.\arabic{equation}}}

\def\endsecteqno{\def{theequation\{\@ifundefined{chapter}%
{\arabic{equation}}{\thechapter.\arabic{equation}}}}

\makeatletter

\newcommand{\s}{\\ \vspace {-3mm} }

\newcommand{\gsim}{\raisebox{-0.13cm}{~\shortstack{$>$ \\[-0.07cm] $\sim$}}~}

\newcommand{\beq}{\begin{eqnarray}}
\newcommand{\eeq}{\end{eqnarray}}
\newcommand{\ba}{\begin{array}}
\newcommand{\ea}{\end{array}}

\def\pl#1#2#3{{\it Phys. Lett. }{\bf B#1~}(19#2)~#3}
\def\zp#1#2#3{{\it Z. Phys. }{\bf C#1~}(19#2)~#3}
\def\prl#1#2#3{{\it Phys. Rev. Lett. }{\bf #1~}(19#2)~#3}

\def\pr#1#2#3{{\it Phys. Rev. }{\bf D#1~}(19#2)~#3}
\def\np#1#2#3{{\it Nucl. Phys. }{\bf B#1~}(19#2)~#3}

\begin{document}
\setlength{\unitlength}{1cm}
%\secteqno

%

\begin{titlepage}

\begin{flushright}
KA--TP--11--1997\\
DFPD/97/28 \\
{\tt hep-ph/9707437}\\
July 1997 \\
\end{flushright}

\def\thefootnote{\fnsymbol{footnote}}

\vspace{1cm}

\begin{center}

{\large\sc {\bf One--loop MSSM Contribution to the Weak Magnetic}}

\vspace{.3cm}

{\large\sc {\bf Dipole Moments of Heavy Fermions}} 

\vspace{1cm}

{\sc W.~Hollik${}^{a}$,
J.I.~Illana${}^a$,
S.~Rigolin${}^{a,b}$,
D.~St\"ockinger${}^{a}$} 
\footnote{E-mail addresses: 
\{hollik,jillana,ds\}@itpaxp3.physik.uni-karlsruhe.de, rigolin@pd.infn.it}  

\vspace{1cm}
${}^a$
Institut f\"ur Theoretische Physik, Universit\"at Karlsruhe,\\
D--76128 Karlsruhe, FR Germany 
\\\vspace{.3cm}
${}^b$
Dipartimento di Fisica, Universit\`a di Padova and INFN,\\
I-35131 Padua, Italy \\

\end{center}

\vspace{1cm}

\begin{abstract}

The MSSM predictions at the one-loop level for the weak-magnetic dipole moments
of the $\tau$ lepton and the $b$ quark are analysed. The entire 
supersymmetric parameter space is scanned with the usual GUT constraint and 
common squark and slepton mass parameters. The real part of $a^W_\tau$ is 
dominated by the chargino contribution, being the same order as the SM one or 
even larger in the high $\tan\beta$ region whereas the imaginary part, due to 
Higgs boson diagrams, is negligible compared to the SM value. The real part of
$a^W_b$ is controlled mainly by charginos and also by gluinos, when the
mixing in the bottom squark sector is large, to yield, for high $\tan\beta$,
a contribution one order of magnitude larger than the pure electroweak SM
value but a factor five smaller than the standard QCD contribution.
The imaginary part of $a^W_b$ is the same order as in the SM.

\end{abstract}

\end{titlepage}

\setcounter{page}{2}
\setcounter{footnote}{0}
%
%%%%%%%%%%%%%%%%%%%%%%%%%%%%%%%%%%%%%%%%%%%%%%%%%%%%%%%%%%%%%%%%%%%%%%%%%%%
%
%\section*{Introduction}

The investigation of the electric and magnetic dipole moments 
provides very accurate tests of the quantum structure of the Standard
Model (SM) (see \cite{smemdm} and references therein) and of possible 
extensions (see \cite{neunpemdm,munpemdm} and references therein). 
The fermion-$Z$ boson vertex, in higher order, involves coupling terms 
which in their Lorentz structure are analogous to the magnetic and electric 
dipole terms of the fermion-photon vertex with valuable informations on
the CP-conserving and CP-violating sectors of the SM.
In addition, the magnetic weak dipole terms exhibit non-decoupling properties 
\cite{ber97} when the top quark is involved in the quantum contributions. 
Together with the chirality-flipping character of the dipole form factors 
one thus might expect also some insight into the mechanism of mass generation.
Extensions of the minimal model by new renormalizable interactions
influence the dipole form factors by non-standard loop contributions.
In this article we consider the Minimal Supersymmetric Standard Model (MSSM)
as a particularly interesting extension of the SM and investigate the impact 
of the MSSM one-loop contributions to the weak-magnetic dipole moments of 
heavy fermions.\s

The most general Lorentz structure of the vertex function that 
couples a $Z$ boson and two on-shell fermions (with outgoing momenta
$q$ and $\bar{q}$) can be written as 
\beq
\Gamma^{Zff}_\mu&=&ie\Bigg\{\gamma_\mu\left[\left(
F_{\rm V}-\frac{v_f}{2s_Wc_W}\right)-\left(F_{\rm A}-\frac{a_f}{2s_Wc_W}\right)
\gamma_5\right] \nonumber \\ & & +(q-\bar{q})_\mu[F_{\rm M}+F_{\rm E}\gamma_5] 
-(q+\bar{q})_\mu[F_{\rm S}+F_{\rm P}\gamma_5]\Bigg\}\ ,
\label{vertex}
\eeq
where $v_f\equiv(I^f_3-2s^2_WQ_f)$, $a_f\equiv I^f_3$ (the weak isospin of the
fermion) and $s^2_W\equiv1-c^2_W\equiv\sin^2\theta_W$.
The form factors $F_i(s)$ are defined splitting off the tree level SM terms,  
and depend only on $s\equiv p^2=(q+\bar{q})^2$.
Conservation of the vector current (U(1) gauge invariance) 
constrains the scalar form factor $F_{\rm S}$ to be zero at $s=M^2_Z$. 
The axial form factor $F_{\rm P}(M^2_Z)$ must also vanish for massless fermions.
The vector and axial-vector form factor, $F_{\rm V}$ 
and $F_{\rm A}$, are the only ones that have to be renormalized while the 
others are finite.
The form factors $F_{\rm M}$ and $F_{\rm E}$ are related to the 
weak dipole moments of the fermion $f$ with mass $m_f$ as follows \cite{ber94}:
\beq
{\rm AWMDM}\equiv a^W_f&=&-2m_f\ F_{\rm M}(M^2_Z) \ , \nonumber\\
{\rm WEDM}\equiv d^W_f&=&ie\ F_{\rm E}(M^2_Z) \ .\nonumber
\eeq
The anomalous weak-magnetic dipole moment (AWMDM) is the analogue of the
anomalous magnetic dipole moment (AMDM) $a^\gamma_f=(g_f-2)/2$. 
The $F_{\rm M}$ ($F_{\rm E}$) form factors are the coefficients 
of the {\em chirality-flipping} term of the CP-conserving (CP-violating) 
effective Lagrangian describing $Z$-fermion couplings. Therefore, they 
are expected to get contributions proportional to some positive power of 
the mass of the fermions involved. This allows the construction of observables
which can be probed experimentally most suitably by heavy fermions. Hence,
for on-shell $Z$ bosons, where the dipole form factors are gauge independent,
the $b$ quark and $\tau$ lepton are the most promising candidates.\s

\begin{figure}[htb]
\begin{center}
\begin{picture}(15,12)
\epsfxsize=16cm
\put(-1,0){\epsfbox{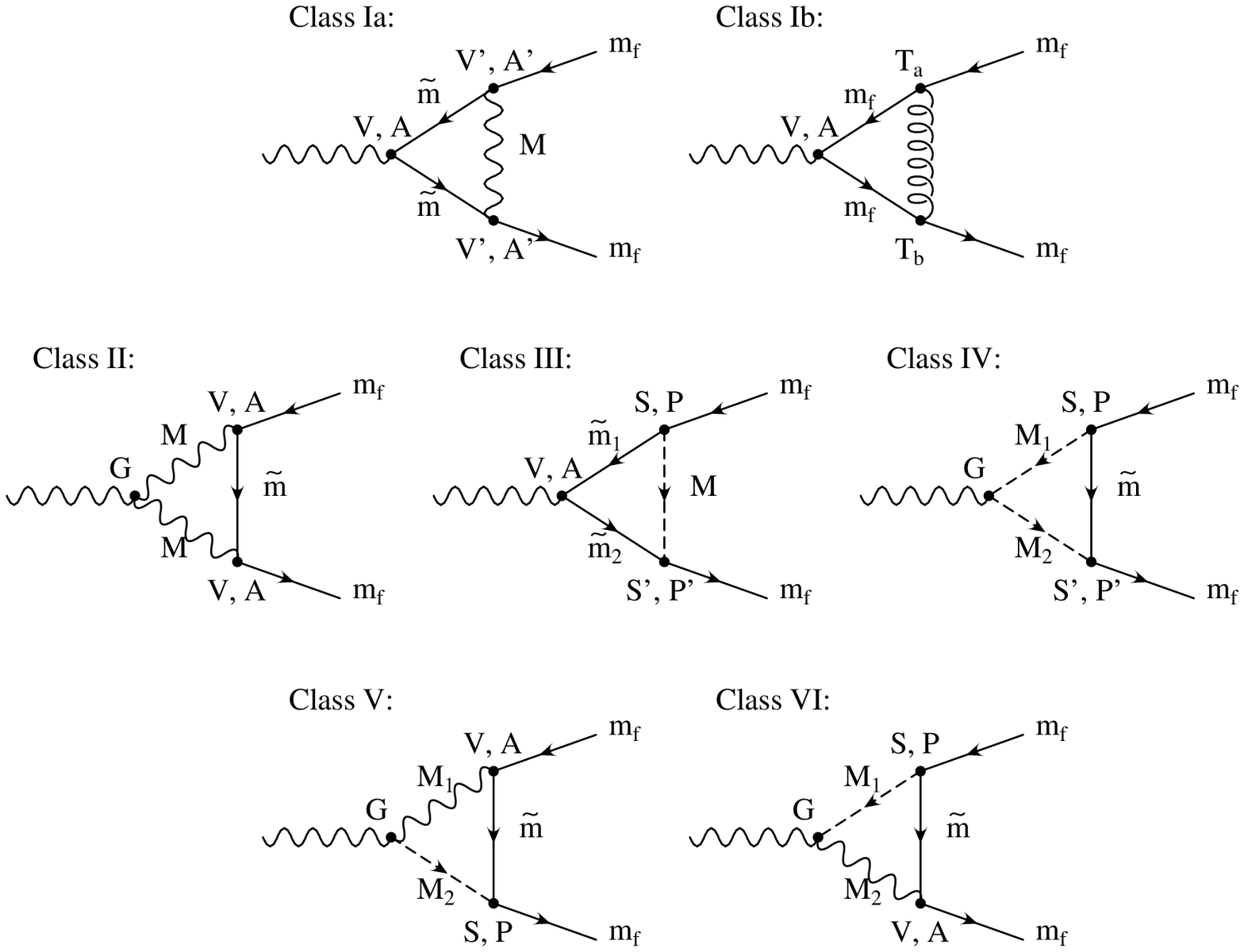}}
\end{picture}
\end{center}
\caption{\sl \small The one-loop $Zff$ diagrams with general couplings.
\label{fig:topo}}
\end{figure}

The contribution to the AWMDM, for the $\tau$  lepton and for heavy quarks, 
at the one-loop level has recently been calculated by Bernab\'eu et al. in the SM 
\cite{ber97,ber95} and in two Higgs  doublet models \cite{ber95b} . 
In this letter, we extend these calculations to the MSSM and investigate the maximum 
size of the non-standard contributions.
We classify all the triangle diagrams in six topologies (classes) or 
generic diagrams (Fig.~\ref{fig:topo}) as in Ref~\cite{topol}. 
Working in the 't Hooft-Feynman gauge all the would-be-Goldstone bosons are 
included.
The global result, adding all the diagrams, is gauge independent.
Every class of diagrams is calculated analytically for general couplings and 
expressed in terms of standard one-loop integrals $\bar{C}$ \cite{topol} and 
vertex coefficients $\lambda$ as follows:

\begin{itemize}

\item\underline{Class Ia}: $\bar{C}=\bar{C}(-\bar{q},q,\tilde{m},\tilde{m},M)$,
\beq
F^{\rm Ia}_{\rm M}(s)&=&\frac{\alpha}{4\pi}\{4m_f\lambda^+_V[-2C^+_2+3C^+_1-C_0]
          +4\tilde{m}\lambda'_V[-2 C^+_1+ C_0]\} \ ,
\eeq
with
$\lambda^\pm_V=V(V'^2+A'^2)\pm 2 AV'A'$, 
$\lambda'_V   =V(V'^2-A'^2)$.
                
\item\underline{Class Ib}: $\bar{C}=\bar{C}(-\bar{q},q,m_f,m_f,0)$
\beq
F^{\rm Ib}_{\rm M}(s)&=&C_{\rm F}\frac{\alpha_s}{4\pi}
        \{V[4m_f(-2 C^+_2+3 C^+_1- C_0)+4\tilde{m} (-2 C^+_1+ C_0)]\} \ .
\eeq
The colour factor $C_{\rm F}=\frac{4}{3}$.
        
\item\underline{Class II}: $\bar{C}=\bar{C}(-\bar{q},q,M,M,\tilde{m})$,
\beq
F^{\rm II}_{\rm M}(s)&=&\frac{\alpha}{4\pi}\{2m_f\lambda_V[4 C^+_2+ C^+_1]
          -6\tilde{m} \lambda'_V C^+_1\} \ ,
\eeq
with
$\lambda_V  =G(V^2+A^2)$,
$\lambda'_V =G(V^2-A^2)$.

\item\underline{Class III}:
       $\bar{C}=\bar{C}(-\bar{q},q,\tilde{m}_1,\tilde{m}_2,M)$,
\beq
F^{\rm III}_{\rm M}(s)&=&\frac{\alpha}{4\pi}\{-2m_f\lambda^+_V[2 C^+_2-C^+_1]
             +(\tilde{m}_1+\tilde{m}_2)\lambda'_V C^+_1
             +(\tilde{m}_1+\tilde{m}_2)\lambda''_A C^-_1 \nonumber\\ 
      & & \hspace{1cm}+(\tilde{m}_1-\tilde{m}_2)\lambda''_A C^+_1
             +(\tilde{m}_1-\tilde{m}_2)\lambda'_V C^-_1 \} \ ,
\label{class3}
\eeq
with
$\lambda^\pm_V=V(SS'-PP')\pm A(SP'-PS')$,
$\lambda'_V   =V(SS'+PP')$,
$\lambda''_A  =A(SP'+PS')$.

\item\underline{Class IV}: $\bar{C}=\bar{C}(-\bar{q},q,M_1,M_2,\tilde{m})$,
\beq
F^{\rm IV}_{\rm M}(s)&=&\frac{\alpha}{4\pi}\{2m_f\lambda_V[2C^+_2-C^+_1]
             +\tilde{m}\lambda'_V [2C^+_1-C_0]\} \ , 
\label{class4}
\eeq
with
$\lambda_V  =G(SS'-PP')$,
$\lambda'_V =G(SS'+PP')$.

\item\underline{Class V}: $\bar{C}=\bar{C}(-\bar{q},q,M_1,M_2,\tilde{m})$,
\beq
F^{\rm V}_{\rm M}(s)&=&\frac{\alpha}{4\pi}\{ \lambda_V[C^+_1+C^-_1]\} \ ,
\eeq
with
$\lambda_V  =G(SV+PA)$.

\item\underline{Class VI}: $\bar{C}=\bar{C}(-\bar{q},q,M_1,M_2,\tilde{m})$,
\beq
F^{\rm VI}_{\rm M}(s)&=&\frac{\alpha}{4\pi}\{ \lambda_V[C^+_1-C^-_1]\} \ ,
\eeq
with
$\lambda_V  =G(SV-PA)$.
\end{itemize}
We use the conventions of Refs.~\cite{HaberKane} for the 
couplings, masses and vertex coefficients.\footnote{
This leads to the different signs of the tree level part of Eq.~(\ref{vertex})
when compared to Eq.~(3.14) of Ref.~\cite{topol}.}
We employ the compact notation $a^W_f[ijk]$ for
the contribution to the AWMDM of particles $i,j,k$ running in the loop as 
labelled in Fig.~C.1 of Ref.~\cite{topol}. Notice that all the masses 
appearing explicitly as factors of the $\bar{C}$ integrals in the  expressions 
are fermion masses, consistently with the chirality flipping character of the 
dipole moments.

\section*{The WMDM of the \boldmath$\tau$ lepton}

In the SM there are 14 diagrams at one-loop, in the 't Hooft-Feynman gauge, 
contributing to the $\tau$-$Z$ boson coupling.
We are in agreement\footnote{The global sign is due to our different
conventions.} with Ref.~\cite{ber95} with the total value for $M_{H^0}=2M_Z$, 
$m_\tau=1.777$ GeV, $M_Z=91.19$ GeV, $s^2_W=0.232$ and $\alpha=1/128$:
\beq
a^W_\tau[{\rm SM}]=(2.10+0.61\ i)\times10^{-6}\ .
\nonumber
\eeq
The diagrams with Higgs bosons have only a small impact, changing the final 
result \cite{ber95} by less than 1\% for $1<M_{H^0}/M_Z<3$.\s

The MSSM contributions introduce new
parameters, the values of which are not known but are constrained by present
experiments. To reduce the number of free parameters, we  
assume the usual GUT constraint for the soft-breaking mass terms in the 
gaugino sector and a common squark mass parameter as well as a common slepton 
mass parameter. 
For given $\tan\beta$, we are left with the following free parameters:
the gaugino mass parameter for the SU(2) sector $M$,
the Higgs-higgsino mass parameter $\mu$, 
the common slepton soft-breaking mass parameter $m_{\tilde l}$, 
the common squark soft-breaking mass parameter $m_{\tilde q}$, 
the trilinear soft-breaking term $A_\tau$ and
the mass of the pseudoscalar Higgs boson $M_A$. 
We consider two typical scenarios: low and high $\tan\beta$, respectively
$\tan\beta=1.6$ and 50.
The genuine MSSM diagrams to be included are:
diagrams with MSSM Higgs bosons; diagrams with charginos and scalar 
neutrinos; and diagrams with neutralinos and $\tilde\tau$ sleptons.

\subsubsection*{MSSM Higgs contribution to \boldmath$a^W_\tau$}

The diagrams for $Z\tau\tau$ involving Higgs bosons are:
\begin{description}
\item
class III: [$\tau\tau h$], [$\tau\tau H$], [$\tau\tau A$], [$\tau\tau G^0$], 
           [$\nu\nu H^-$], $[\nu\nu G^-$],
\item
class IV: [$Ah\tau$], [$G^0h\tau$], [$AH\tau$], [$G^0H\tau$],
          [$hA\tau$], [$hG^0\tau$], [$HA\tau$], [$HG^0\tau$],
          [$H^-H^-\nu$], [$H^-G^-\nu$], [$G^-H^-\nu$], [$G^-G^-\nu$], 
\item
class V: [$Zh\nu$], [$ZH\nu$], [$WG^-\nu$],
\item
class VI: [$hZ\nu$], [$HZ\nu$], [$G^-W\nu$],
\end{description}
\noindent
where $G^0$, $G^\pm$ are the would-be-Goldstone bosons and $h$, $H$, $A$ and 
$H^\pm$ are the physical MSSM Higgs bosons.
Actually, not all of them give a contribution to the AWMDM, in particular, the 
diagrams of class IV with neutral Higgs bosons vanish as in the case of the SM.
The masses of the Higgs bosons are fixed by the mass of the pseudoscalar,
$M_A$, $\tan\beta$ and the common squark mass parameter $m_{\tilde q}$,
which are the only free parameters involved in this calculation. A value
$m_{\tilde q}=250$ GeV will be assumed in the following. Low values
of $M_A$ yield the largest contribution and for $M_A\gsim$ 300 GeV the 
contributions become $M_A$-independent.
The class III diagrams provide the only contribution to Im($a^W_\tau$),
assuming the present lower bound for the masses of the Higgs bosons 
\cite{higgs}.
Due to the fact that the ratio of vector couplings $v_\tau/v_\nu$ is small, 
the diagrams with charged Higgs bosons give the main contribution,
which is of the order $\alpha/4\pi(m_\tau/M_Z)^4\ 
\tan^2\beta \sim 10^{-10}\ (10^{-7})$ for the low (high) $\tan\beta$
scenario. For example, one gets
\beq
{\rm Im}(a^W_\tau[{\rm Higgs}])=-0.0001\ (0.06)\times10^{-6} 
\mbox{ for } M_A=100 \mbox{ GeV and low (high)} \tan\beta\ .
\nonumber
\eeq
For Re($a^W_\tau$), the diagrams of classes III and IV are proportional
to $(m_\tau/M_Z)^4\tan^2\beta$ and the diagrams of classes V and VI are
proportional to $(m_\tau/M_Z)^2$. Actually,
\beq
{\rm Re}(a^W_\tau[{\rm Higgs}])=\left\{\ba{r}
-0.3\ (-0.4)\times 10^{-6} \mbox{ for }M_A=100 \mbox{ GeV} \\
-0.3\ (-0.3)\times 10^{-6} \mbox{ for }M_A\gsim300\mbox{ GeV} \ea\right.
\mbox{ for low (high) }\tan\beta\ .  
\nonumber
\eeq

\subsubsection*{Chargino and scalar neutrino contribution to 
\boldmath$a^W_\tau$}

There are two diagrams involving charginos and scalar neutrinos: 
\begin{description}
\item
class III: [$\tilde{\chi}^-_i\tilde{\chi}^-_j\tilde\nu$],
\item
class IV: [$\tilde\nu\tilde\nu\tilde{\chi}^-_k$].
\end{description}
Charginos \cite{neutracharg} and scalar neutrinos \cite{sneutrinos} do not 
occur in $Z$ decays and hence their contribution to the AWMDM is real, in the
region of their masses not ruled out by the experiments.
The free parameters involved here are $\tan\beta$, $M$, $\mu$ and 
$m_{\tilde l}$. The contribution becomes smaller
increasing $m_{\tilde l}$ (consistently with decoupling \cite{decoupling}). 
In the large $M$ and $|\mu|$ region the charginos also decouple.
The contribution is enhanced by increasing $\tan\beta$.
In the low $\tan\beta$ scenario the contribution is of order $10^{-7}$.
The value for $m_{\tilde l}=250$ GeV, $M=200$ GeV and $|\mu|=200$ GeV
is
\beq
{\rm Re}(a^W_\tau[{\rm charginos}])= \left\{ \ba{r}
-0.2\times10^{-6}\ (\mu<0) \\ +0.2\times10^{-6}\ (\mu>0)
\ea\right.\ \mbox{ for low}\ 
\tan\beta\ .
\nonumber
\eeq
In the high $\tan\beta$ scenario the contribution is of order $10^{-5}$.
We find:
\beq
{\rm Re}(a^W_\tau[{\rm charginos}])= \left\{ \ba{r}
-7.0\times10^{-6}\ (\mu<0) \\ +7.0\times10^{-6}\ (\mu>0)
\ea\right.\  
\mbox{ for high}\ \tan\beta
\nonumber
\eeq 
and the same values for $M$ and $|\mu|$ as above. This enhancement for higher 
$\tan\beta$ is expected in analogy to the AMDM of the muon. 
The contours in the $M-\mu$ plane are similar to the ones for the $b$ quark 
(Fig.~\ref{fig:botchar}) suppressed by factors between $m_\tau/m_b$ and
$(m_\tau/m_b)^2$ [see Eqs.~(\ref{class3}) and (\ref{class4})].

\subsubsection*{Neutralino and \boldmath$\tilde\tau$ slepton contribution to 
\boldmath$a^W_\tau$}

There are two diagrams involving neutralinos and $\tilde\tau$ sleptons: 
\begin{description}
\item
class III: [$\tilde{\chi}^0_i\tilde{\chi}^0_j\tilde\tau_k$],
\item
class IV: [$\tilde\tau_i\tilde\tau_j\tilde{\chi}^0_k$].
\end{description}
The $\tilde\tau$ slepton masses are bounded by present experiments \cite{sleptons}
to be heavier than half the mass of the $Z$, but the neutralinos are less 
stringently bounded and could be much lighter \cite{neutracharg}, allowing for 
the possibility of a contribution to the imaginary part of the AWMDM through 
diagrams of class III.
Now, we have $\tan\beta$, $M$, $\mu$, $m_{\tilde l}$, as well as the
trilinear soft-breaking mass term for the $\tilde\tau$ sleptons, $A_\tau$, as 
free parameters.
We choose different values for $A_\tau$ in the range $\pm\mu\tan\beta$ to 
estimate the sensitivity of the AWMDM to this soft-breaking term and find no 
sizeable deviation. However, for some values of $A_\tau$ and
$\mu$ the mass of the $\tilde\tau$ sleptons can become unphysical. For 
simplicity, we take the value of
$A_\tau$ that makes $m^\tau_{LR}=A_\tau-\mu\tan\beta=0$.
Just below the neutralino threshold the imaginary part is
\beq
|{\rm Im}(a^W_\tau[{\rm neutralinos}])|\sim 10^{-9}\ (10^{-7}) \mbox{ for 
low (high)}\ \tan\beta
\nonumber
\eeq
and it is negligible far below threshold. 
Concerning the real part, also the value is maximum near the neutralino
pair threshold.
Like in the chargino-scalar neutrino contributions, the result
is expected to be enhanced by $\tan\beta$. We find
\beq
{\rm Re}(a^W_\tau[{\rm neutralinos}])=-0.02(-0.4)\times10^{-6} \mbox{ for 
low (high)}\ \tan\beta
\nonumber
\eeq
and $M=\mu=200$ GeV.

\section*{The WMDM of the \boldmath$b$ quark}

In the electroweak SM there are 14 diagrams, in the 't Hooft-Feynman gauge,
contributing to the $b$-$Z$ boson coupling at the one-loop level, plus one 
more (gluon exchange) if QCD is also included. 
Our calculation reproduces the result of Ref.~\cite{ber97}.
Taking as input  $m_b=4.5$ GeV, $m_t=175$ GeV, 
$M_Z=91.19$ GeV, $s^2_W=0.232$, $\alpha=1/128$ and $\alpha_s=0.118$,
the pure electroweak contribution is  
$a^W_b({\rm EW})=[(1.1;\ 2.0;\ 2.4)-0.2\ i]\times10^{-6}$,
for $M_{H^0}=M_Z,\ 2M_Z,\ 3M_Z$,
but the whole value is dominated by the QCD contribution to give
\beq
a^W_b[{\rm EW+QCD}]=(-2.98+1.56\ i)\times10^{-4}\ .
\nonumber
\eeq
Only the third generation of quarks is considered and the element $V_{tb}$ 
of the CKM matrix is set to 1. \s 

In the MSSM one has to include the following genuine supersymmetric 
contributions: diagrams with MSSM Higgs bosons; diagrams with charginos and
$\tilde t$ squarks; diagrams with neutralinos and $\tilde b$ squarks; and diagrams with 
gluinos and $\tilde b$ squarks.
In this case, we are left with the following free parameters:
$M$, $\mu$, $m_{\tilde q}$, $M_A$,
the trilinear soft-breaking terms $A_b$ and $A_t$ and
a mass for the gluinos $m_{\tilde g}$.

\subsubsection*{MSSM Higgs contribution to \boldmath$a^W_b$}

The diagrams for $Zbb$ involving Higgs bosons are
\begin{description}
\item
class III: [$bbh$], [$bbH$], [$bbA$], [$bbG^0$], 
           [$ttH^-$], [$ttG^-$],
\item
class IV: [$Ahb$], [$G^0hb$], [$AHb$], [$G^0Hb$],
          [$hAb$], [$hG^0b$], [$HAb$], [$HG^0b$],
          [$H^-H^-t$], [$H^-G^-t$], [$G^-H^-t$], [$G^-G^-t$], 
\item
class V: [$Zht$], [$ZHt$], [$WG^-t$],
\item
class VI: [$hZt$], [$HZt$], [$G^-Wt$].
\end{description}
As before, the diagrams of class IV with neutral Higgs bosons do not
contribute.
The diagrams of class III with neutral 
Higgs bosons yield Im$(a^W_b)\sim \alpha/4\pi(m_b/M_Z)^4\tan^2\beta \sim 10^{-8}
\ (10^{-6})$ for low (high) $\tan\beta$. Taking $m_{\tilde q}=250$ GeV
we get
\beq
{\rm Im}(a^W_b[{\rm Higgs}])=\left\{\ba{r}
0.02\ (3.7)\times10^{-6} \mbox{ for } M_A=100 \mbox{ GeV} \\
0.01\ (0.08)\times 10^{-6} \mbox{ for } M_A=300 \mbox{ GeV} \ea
\right.
\mbox{ and low (high) } \tan\beta\ .
\nonumber
\eeq
Concerning Re($a^W_b$),
there are contributions proportional to 
$\displaystyle\left(\frac{m_b}{M_Z}\right)^2$, 
$\displaystyle\left(\frac{m_bm_t}{M_Z^2}\right)^2$
and $\displaystyle\left(\frac{m_b}{M_Z}\right)^4\tan^2\beta$.
%$(m_b/M_Z)^2$, $(m_bm_t/M_Z^2)^2$ and $(m_b/M_Z)^4\tan^2\beta$. 
The sum is thus not very
sensitive to $\tan\beta$,
\beq
{\rm Re}(a^W_b[{\rm Higgs}])= \left\{\ba{r}
-3.8\ (+0.8)\times10^{-6} \mbox{ for } M_A=100 \mbox{ GeV} \\
-1.8\ (-0.9)\times 10^{-6} \mbox{ for } M_A=300 \mbox{ GeV} \ea
\right.
\mbox{ and low (high) } \tan\beta\ .
\nonumber
\eeq

\subsubsection*{Chargino and \boldmath$\tilde t$ squark contribution to 
\boldmath$a^W_b$}

\begin{figure}[htb]
\begin{center}
\begin{picture}(15,6.75)
\epsfxsize=8cm
\put(-1,-0.75){\epsfbox{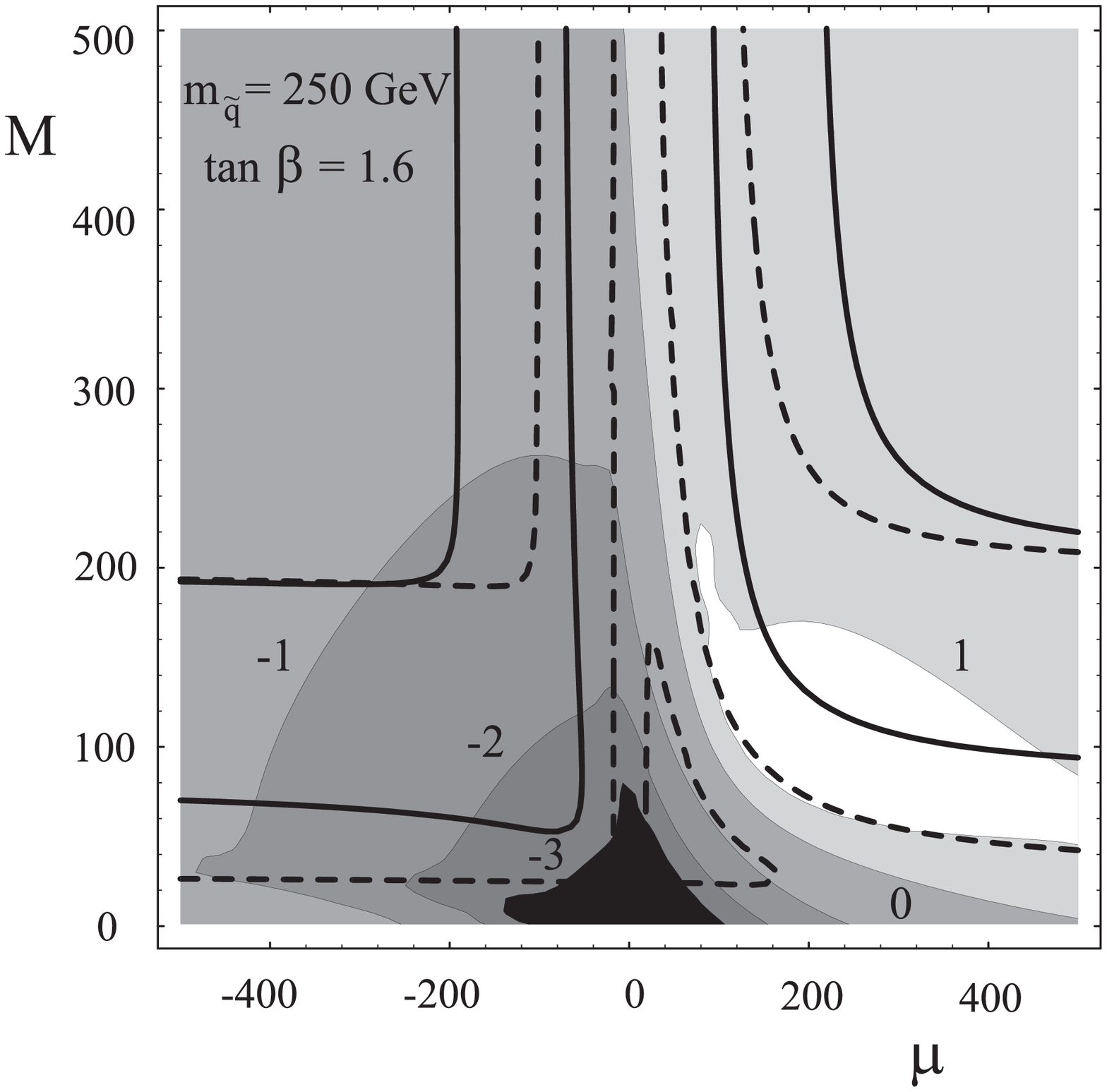}}
\epsfxsize=8cm
\put(7.5,-0.75){\epsfbox{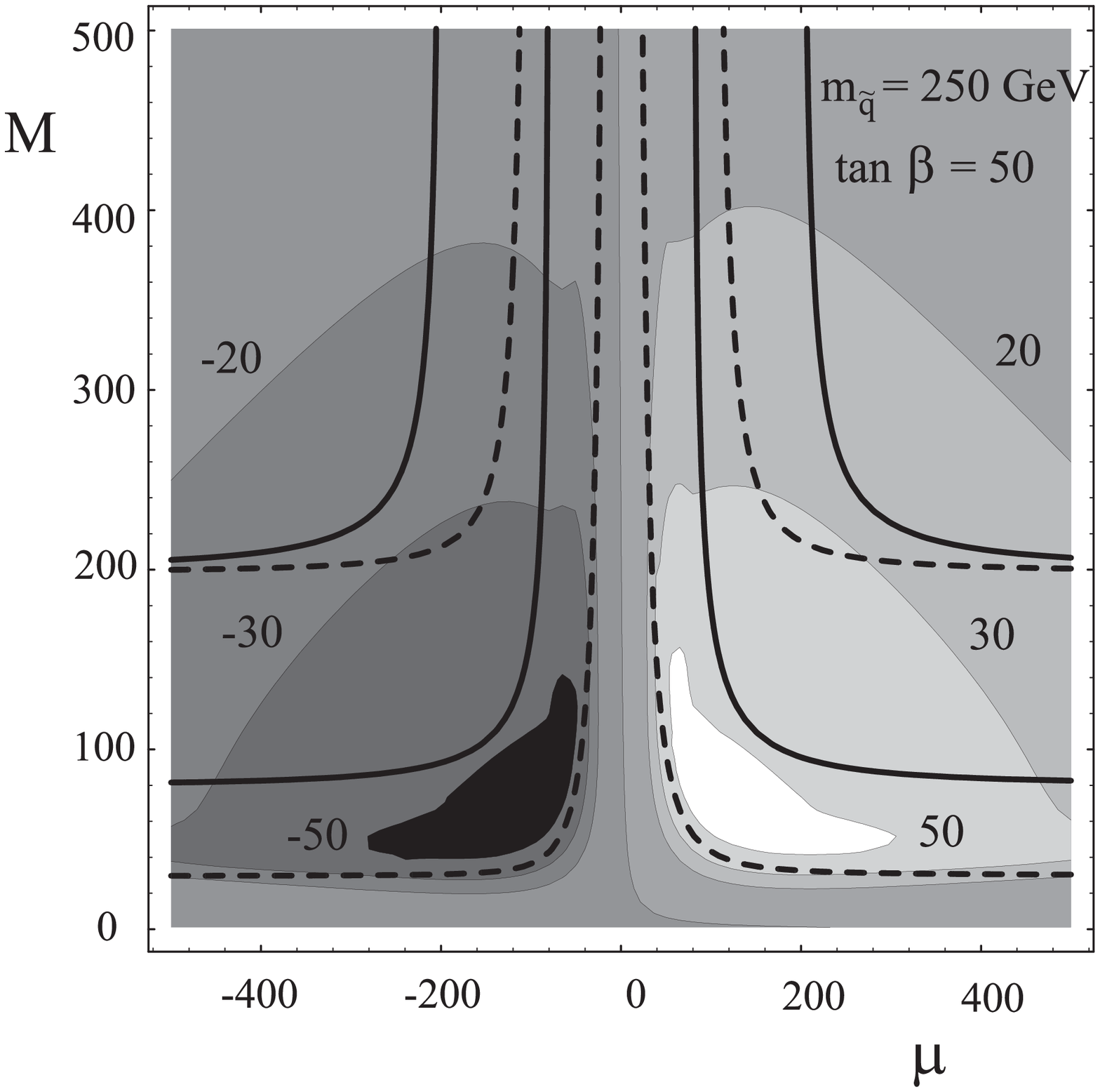}}
\put( 3.35,-0.7){(a)}
\put(11.85,-0.7){(b)}
\end{picture}
\end{center}
\caption{\sl \small Re($a^W_b$[charginos]) for low (a) and high (b)
$\tan\beta$ in units of $10^{-6}$ in the plane $M-\mu$. The contour 
solid-lines correspond to the light chargino masses $m_{\tilde\chi^\pm}$=80 and 
200 GeV and the contour dashed-lines correspond to the lightest neutralino 
masses $m_{\tilde\chi^0}$=15 and 100 GeV. The common squark mass parameter is 
fixed to $m_{\tilde q}=250$ GeV and $m^t_{LR}=0$.
\label{fig:botchar}}
\end{figure}

There are two diagrams involving charginos and $\tilde t$ squarks:
\begin{description}
\item
class III: [$\tilde{\chi}^-_i\tilde{\chi}^-_j\tilde{t}_k$],
\item
class IV: [$\tilde{t}_i\tilde{t}_j\tilde{\chi}^-_k$].
\end{description}
As in the $\tau$ case, assuming the present bounds on chargino and squark
masses \cite{neutracharg,squarks-gluinos}, Im$(a^W_b$[charginos])=0.
The free parameters involved are $\tan\beta$, $M$, $\mu$, $m_{\tilde q}$ and
$A_t$.
Varying the value for $A_t$ in the range $\pm\mu\cot\beta$ shows that 
there is only a small effect in the AWMDM. We thus choose $A_t$ in a way that
the off-diagonal entry of the $\tilde{t}$ squarks mixing mass matrix 
$m^t_{LR}=A_t-\mu\cot\beta=0$. Increasing $m_{\tilde q}$ the contribution 
becomes smaller.
Fig.~\ref{fig:botchar} shows a contour plot in the $M-\mu$ plane for 
$m_{\tilde q}=250$ GeV. In the
large $M$ and $|\mu|$ region, the charginos decouple, as expected
from the decoupling theorem \cite{decoupling}.
The contour lines for the masses of the lightest chargino and lightest
neutralino in the $M-\mu$ plane are also depicted. The contributions are
enhanced by increasing $\tan\beta$.
In the low $\tan\beta$ scenario the contribution is of order $10^{-6}$
(Fig.~\ref{fig:botchar}a). For $M=|\mu|$=200 GeV we get
\beq
{\rm Re}(a^W_b[{\rm charginos}])=\left\{ \ba{r}
-1.1\times10^{-6}\ (\mu<0) \\ 1.0\times10^{-6}\ (\mu>0)
\ea\right.\ \mbox{ for low}\ \tan\beta\ .
\nonumber
\eeq
In the high $\tan\beta$ scenario (Fig.~\ref{fig:botchar}b) the contribution 
is enhanced to become of order $10^{-5}$. For the same values of $M$ and $\mu$
as above,  
\beq
{\rm Re}(a^W_b[{\rm charginos}])\simeq \left\{ \ba{r}
-32.2\times10^{-6}\ (\mu<0) \\ +33.6\times10^{-6}\ (\mu>0)
\ea\right.\  
\mbox{ for high}\ \tan\beta\ .
\nonumber
\eeq 

\subsubsection*{Neutralino and \boldmath$\tilde b$ squark contribution to 
\boldmath$a^W_b$}

There are two diagrams involving neutralinos and $\tilde b$ squarks:
\begin{description}
\item
class III: [$\tilde{\chi}^0_i\tilde{\chi}^0_j\tilde{b}_k$],
\item
class IV: [$\tilde{b}_i\tilde{b}_j\tilde{\chi}^0_k$].
\end{description}
Now, we have $\tan\beta$, $M$, $\mu$, $m_{\tilde q}$, as well as the
trilinear soft-breaking mass term for the $\tilde b$ squarks, $A_b$, as free
parameters. There is no sizeable dependence on $A_b$. 
As in the $\tau$ case, due to the heavy mass of the $\tilde b$ squarks
\cite{squarks-gluinos}, an imaginary part can only arise 
through diagrams of class III.
The region of non vanishing Im($a^W_b$) is below the neutralino threshold and 
the value is 
\beq
|{\rm Im}(a^W_b[{\rm neutralinos}])|\sim 0.1 (1) \times 10^{-6} \mbox{ for low 
(high)}\ \tan\beta\ ,
\nonumber
\eeq
near threshold. 
Concerning the real part, also the value is maximum near the neutralino
pair threshold.
The result is proportional to $\tan\beta$ as expected:
\beq
{\rm Re}(a^W_b[{\rm neutralinos}])=-0.2\ (-10.5)\times10^{-6}\  \mbox{ for low 
(high)}\ \tan\beta\ 
\nonumber
\eeq
and $M=\mu=200$ GeV.

\subsubsection*{Gluino contribution to \boldmath$a^W_b$}

There is one diagram with a gluino and two $\tilde b$ squarks running in the 
loop:
\begin{description}
\item
class IV: $[\tilde{b}_i\tilde{b}_j\tilde{g}]$.
\end{description}
The present bounds on $\tilde b$ squark masses \cite{squarks-gluinos} only 
allow a real contribution to the AWMDM.
The free parameters are $\tan\beta$, $m_{\tilde q}$, $m_{\tilde g}$,
$A_b$ and $\mu$. These last two parameters
affect the result only through the off-diagonal term of the
$\tilde b$ squark mass matrix $m^b_{LR}=A_b-\mu\tan\beta$. 
There still remains a slight dependence 
on $\tan\beta$ in the diagonal terms of the $\tilde b$ squark mass matrix and 
so we keep the distinction between low and high $\tan\beta$ scenarios.
The mixing in the $\tilde b$ sector is determined by $m^b_{LR}$ and intervenes
in the contribution due to chirality flipping in
the gluino internal line (the contribution proportional to $m_{\tilde g}$). 
Thus, the contribution to the AWMDM is enhanced by the largest values of 
$m^b_{LR}$ compatible with a physical and experimentally not excluded mass 
for the lightest $\tilde b$ squark. 
Assuming a typical scale for $m^b_{LR}$ of order $\mu\tan\beta$ leads to
values of the order 100 GeV (10 TeV) for the low (high) 
$\tan\beta$ scenarios. \s

\begin{figure}[htb]
\begin{center}
\begin{picture}(15,6.75)
\epsfxsize=8cm
\put(-1,-0.75){\epsfbox{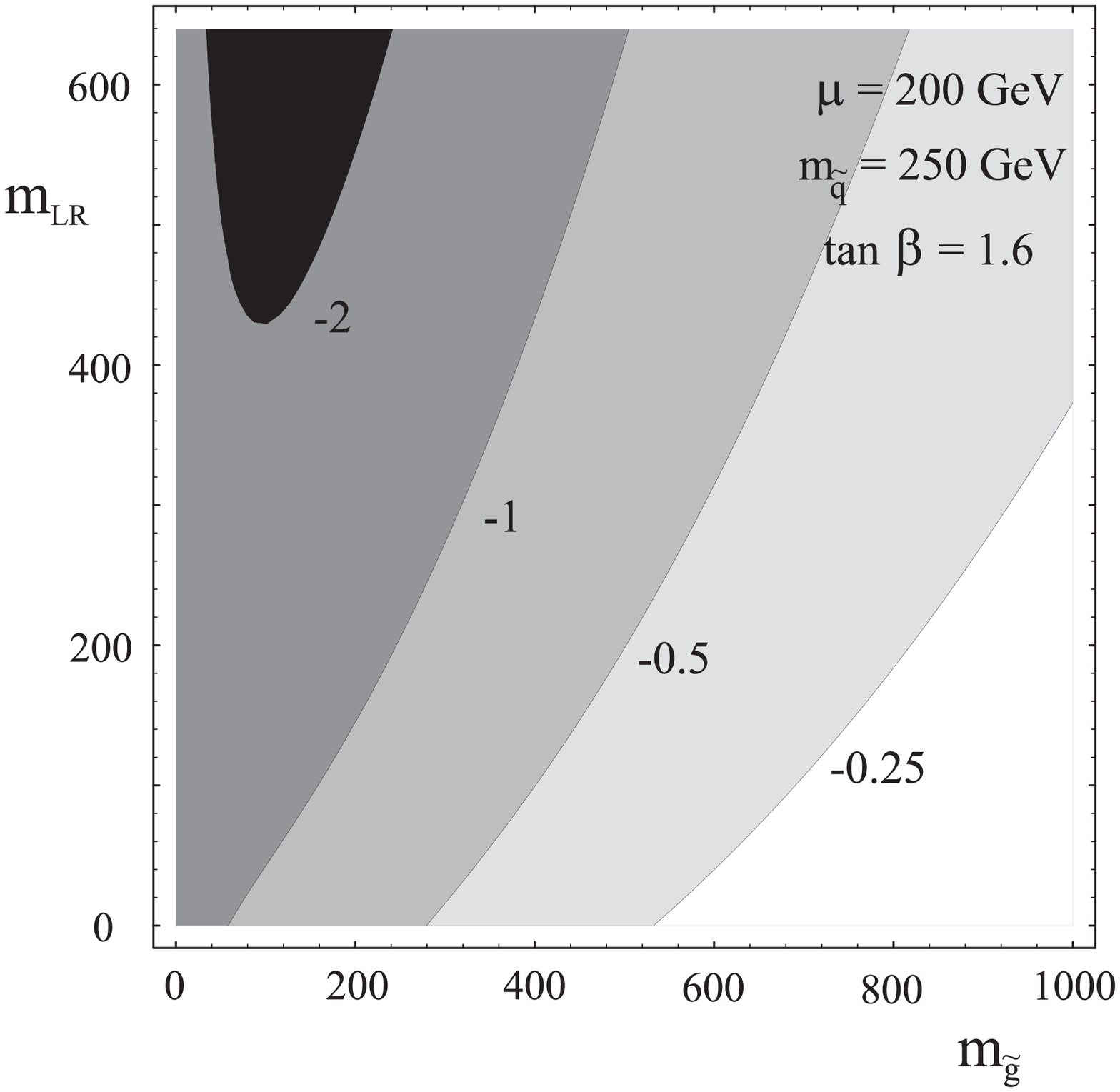}}
\epsfxsize=8cm
\put(7.5,-0.75){\epsfbox{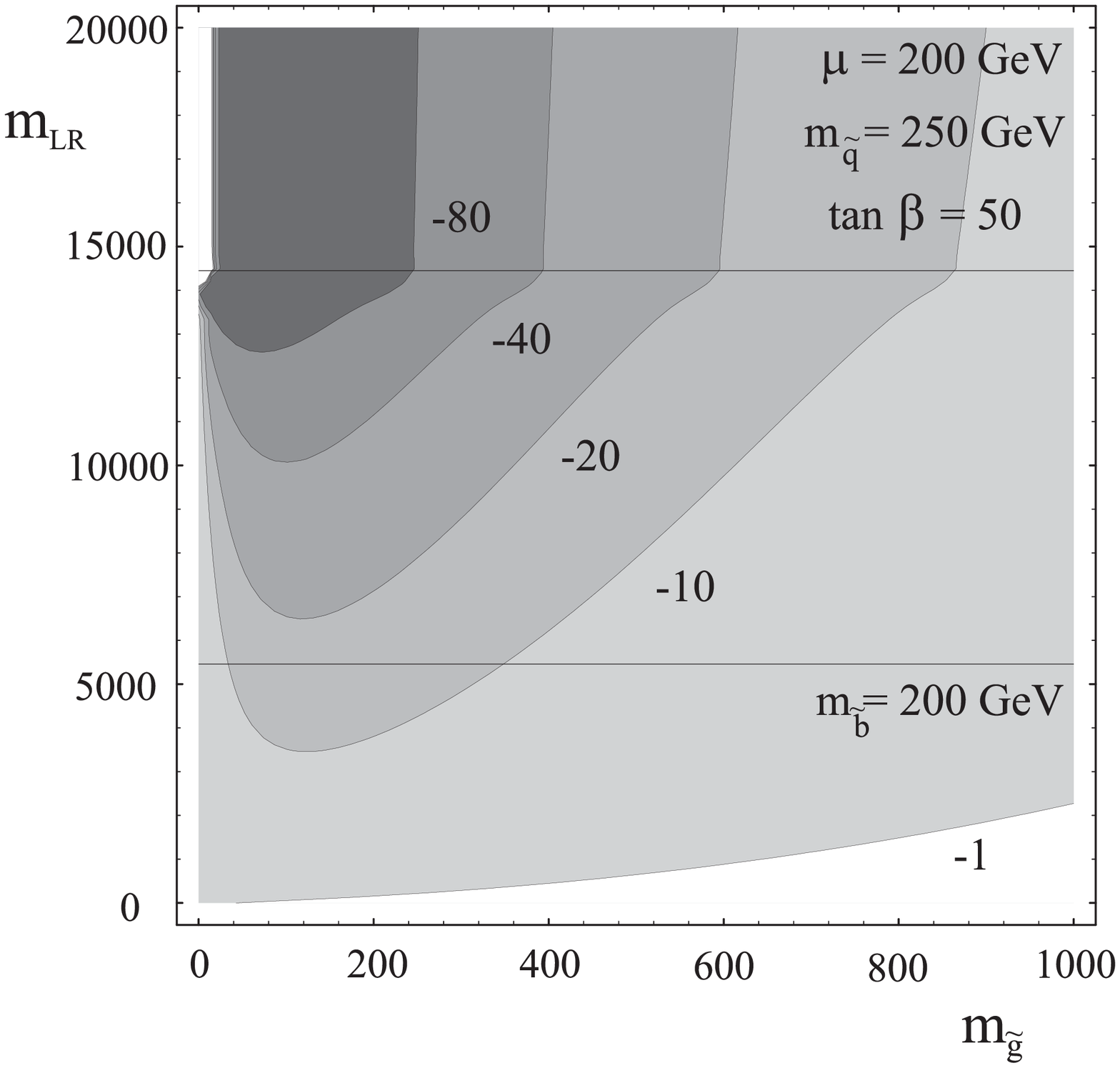}}
\put( 3.35,-0.7){(a)}
\put(11.85,-0.7){(b)}
\end{picture}
\end{center}
\caption{\sl \small Re($a^W_b$[gluinos]) low (a) and high (b) $\tan\beta$ 
in units of $10^{-6}$ in the plane $m^b_{LR}-m_{\tilde g}$. The common squark 
mass parameter is fixed to $m_{\tilde q}=250$ GeV. The region above the upper horizontal 
line is unphysical ($m^2_{\tilde b_1}<0$).\label{fig:gluinos}}
\end{figure}

The low $\tan\beta$ region is displayed in Fig.~\ref{fig:gluinos}a, for
$m_{\tilde q}=250$ GeV. The mixing for the plotted values of $m^b_{LR}$ do
not affect significantly the mass of the lightest $\tilde b$ squark (always
above 200 GeV). For zero gluino mass, only the term proportional to the mass
of the $b$ quark provides a contribution. As we increase the gluino mass, 
the term proportional to $m_{\tilde g}$ dominates, especially for large 
$m^b_{LR}$, being again suppressed at high $m_{\tilde g}$ due to the gluino 
decoupling. The typical value is
\beq
{\rm Re}(a^W_b[{\rm gluinos}])=-1.1 \times 10^{-6}\ \mbox{ for low}\ \tan\beta\ 
\nonumber
\eeq
and $m_{\tilde g}=m^b_{LR}=200$ GeV.
For high $\tan\beta$ (Fig.~\ref{fig:gluinos}b), the behaviour is analogous
but larger values can be obtained. For example, choosing $m_{\tilde g}=200$ GeV 
and $m^b_{LR}=7$ TeV, we obtain
\beq
{\rm Re}(a^W_b[{\rm gluinos}])=-18.9\times10^{-6}\ \mbox{ for high}\ 
\tan\beta\ .
\nonumber
\eeq

\section*{Conclusions}

From the previous analysis we conclude that the imaginary part of
the AWMDM of the $\tau$ lepton is provided only by the diagrams with Higgs
bosons, being at most of order $10^{-7}$ for the high $\tan\beta$ scenario, an 
order of magnitude below the SM contribution.
For the real part, the charginos dominate with a contribution of
order $10^{-5}$ ($10^{-6}$) in the high (low) $\tan\beta$ scenario at
the level of the SM contribution or even larger. The
neutralino contribution is of opposite sign, but one order of magnitude
smaller. Also the Higgs contribution to the real part is negligible. \s

The most important MSSM contribution to the $a^W_b$ is provided by charginos 
and gluinos in the high $\tan\beta$ scenario.
For high $\tan\beta$, the neutralino and chargino contributions have
opposite sign, but the former is typically a factor five below. On the other
hand, the chargino contribution has the sign of $\mu$ whereas the gluinos
only contribute with negative sign. 
Therefore the total contribution to the real part is maximal for negative
$\mu$ and can reach the value of 
\beq
{\rm Re}(a^W_b[{\rm MSSM}])\sim-60\times10^{-6}\ \mbox{ for high}\ \tan\beta\ .
\nonumber
\eeq
These values are one order of magnitude higher than the pure electroweak
SM contribution, but still a factor five below the standard QCD contribution.
The Higgs boson diagrams are negligible,
and they are only important for the imaginary part, contributing the same
amount as the neutralinos below threshold, at most
\beq
|{\rm Im}(a^W_b[{\rm MSSM}])|\sim 10^{-6}\ \mbox{ for high}\ \tan\beta\ . 
\nonumber
\eeq
This value is of the same order as the electroweak SM result.
In the low $\tan\beta$ scenario, all the MSSM contributions to the real
part are of order $10^{-6}$; for the imaginary part they are one order of 
magnitude smaller. They both are comparable in size to the pure electroweak SM
contribution.

\section*{Acknowledgements}

Discussions with  F. Feruglio, G.A. Gonz\'alez-Sprinberg and A. Masiero
are gratefully acknowledged. J.I.I. is supported by the Fundaci\'on Ram\'on
Areces and partially by the Spanish CICYT under contract AEN96-1672. S.R. is 
supported by the Fondazione Ing. A. Gini and by the italian MURST.

%
%%%%%%%%%%%%%%%%%%%%%%%%%%%%%%%%%%%%%%%%%%%%%%%%%%%%%%%%%%%%%%%%%%%%%%%%%%%
%

%
%%%%%%%%%%%%%%%%%%%%%%%%%%%%%%%%%%%%%%%%%%%%%%%%%%%%%%%%%%%%%%%%%%%%%%%%%%%
%
\newpage
\section*{Figures}
\vspace{1cm}

%\begin{figure}[htb]
%\begin{center}
%\begin{picture}(15,7.5)
%\epsfxsize=8cm
%\put(-1,-0.5){\epsfbox{tauclo250.eps}}
%\epsfxsize=8cm
%\put(7.5,-0.5){\epsfbox{tauchi250.eps}}
%\end{picture}
%\end{center}
%\caption{\sl .
%\label{fig:tauchar}}
%\end{figure}

\end{document}